\documentclass{article}
\usepackage{graphicx,epsfig,array}
\usepackage{amssymb,amsmath}
\usepackage{cite}

\begin{document}
   
\begin{center}
\large \bf

On the dynamics of solitary wave solutions supported by
the model of mutually penetrating continua

\end{center}
\vspace*{3 mm}
%
% FULL AUTHOR NAMES, i.e. John Thomas Smith - Capitalize only first letter
%
\centerline{\bf Sergii Skurativskyi \footnote{ Corresponding author: \texttt{skurserg@gmail.com}}
, Vjacheslav Danylenko}

\vspace*{3 mm}

\begin{center}
\textit{Subbotin Institute of Geophysics, NAS of Ukraine, 
Bohdan Khmelnytskyi str. 63-G, Kyiv, UKRAINE}
\end{center}

\vspace*{3 mm}
%
% ABSTRACT goes below.
% Do not delete the {\em Abstract:} word.
%
\textit{ Abstract:} The model we deal with is the mathematical model for mutually penetrating
continua one of which is the carrying medium obeying the wave equation whereas the
other one is the oscillating inclusion described by the equation for oscillators. These
equations of motion are closed by the cubic constitutive equation for the carrying
medium. Studying the wave solutions we reduce this model to a plane dynamical
system of Hamiltonian type. This allows us to derive the relation describing the
homoclinic trajectory going through the origin and obtain the solitary wave with
infinite support. Moreover, there exist a limiting solitary wave with finite support, i.e.
compacton. To model the solitary waves dynamics, we construct the three level
finite-difference numerical scheme and study its stability. We are interested in the
interaction of the pair of solitary waves. It turns out that the collisions of solitary
waves have non-elastic character but the shapes of waves after collisions are
preserved.

%PACS numbers: 89.75.-k, 05.45.Yv, 05.45.-a

\section{Introduction}

Natural geomaterials are highly heterogeneous and interact intensively with the environment. In these conditions the peculiarities of internal medium's structure, namely discreteness and oscillating dynamics \cite{sadov,monog}, can manifest. To incorporate these features of media, the    mathematical model for mutually penetrating
continua is used \cite{Palmov,Slepjan}. This model consists of the wave equation for carrying medium and equations of motion for oscillating continuum which is regarded as the set of partial oscillators. To generalize the linear counterpart of this model  \cite{Palmov,Slepjan}, the nonlinearity has been incorporated in the equation of state for carring medium \cite{nasu,nonlin_dym,Skur_JMS_2014} and in the kinetics for oscillator's equations of motion. We thus are going to treat  the following  mathematical model  
\begin{equation}\label{model}
\rho \frac{{\partial ^2 u}}{{\partial t^2 }} = \frac{{\partial
\sigma }}{{\partial x}} - m \rho  \frac{{\partial ^2 w
}}{{\partial t^2 }}, \qquad \frac{{\partial ^2 w }}{{\partial t^2
}} + \Phi\left( {w - u} \right) = 0,
\end{equation}
where $\rho$ is  medium's density, $u$ and $w$ are the displacements of carrying medium and oscillator from the rest state,  $m\rho$ is the density of oscillating continuum. To close model (\ref{model}), we apply the cubic constitutive equation for the carrying medium
  $
 \sigma  = e_1 u_x  + e_3 u_x^3
$  and $ \Phi(x)=\omega ^2 x+\delta x^3$. The novelty of this model lies in the taking into account the cubic terms in the expressions for $\sigma$ and $\Phi$.

In this report we  consider the properties of  wave solutions having the following form
\begin{equation}\label{solution}
u=U(s), w=W(s), s=x-Dt,
\end{equation}
 where the parameter $D$
 is a  constant velocity of the wave front.
 Inserting (\ref{solution}) into  model (\ref{model}), it easy to see that  the   functions $ U$ and $ W$
 satisfy the dynamical system 
%\begin{equation}
\[
D^2 U' = \rho^{-1}\sigma \left( {U'} \right) - mD^2 W^\prime,
\quad W'' + \Omega ^2 \left( {W - U} \right) +\delta D^{-1}\left(W-U\right)^3= 0,
\]
%\end{equation}
where $\Omega  = \omega D^{-1}$.

This system can be written in the form 
\begin{equation}\label{syst3}
\begin{split}
W' =\alpha _1 R + \alpha _3 R^3, \,
  U' =R, \, (\alpha_1+3\alpha_3 R^2)R'+\Omega^2(W-U)+\delta D^{-2}(W-U)^3=0,
\end{split}
\end{equation}
 where $\displaystyle \alpha _1  =
\frac{{e_1 - D^2 \rho }}{{m \rho D^2 }}$, $\displaystyle \alpha _3
= \frac{{e_3 }}{{m \rho D^2 }}$. Through the report we fix $e_1=1$, $e_3=0.5$, $m=0.6$, $\omega=0.9$ in numerical treatments.

\subsection{Solitary waves in the model with the linear equation of motion for oscillating inclusions}\label{sec11}

At first, consider system (\ref{syst3}) at $\delta=0$. Excluding the variable $W$,  system (\ref{syst3}) is reduced to the planar system 
\begin{equation}\label{syst4}
\begin{split}
  \left(\alpha _1  + 3\alpha _3 R^2 \right)^2 R' = Z \left(\alpha _1  + 3\alpha _3 R^2 \right)^2,\\
 \left(\alpha _1  + 3\alpha _3 R^2 \right)^2 Z' = -\left[ 6\alpha _3 RZ^2  + \Omega ^2 \left( {\left( {\alpha
_1 - 1} \right)R + \alpha _3 R^3 } \right)\right]\left(\alpha _1
+ 3\alpha _3 R^2\right),
\end{split}
\end{equation}
which  admits  the Hamiltonian  
$$
H = \frac{1}{2}Z^2 \left( {\alpha _1  + 3\alpha _3 R^2 } \right)^2
+ \frac{{\Omega ^2 }}{2}\left[ {\alpha _3^2 R^6  + \frac{{4\alpha
_1 - 3}}{2}\alpha _3 R^4  + \left( {\alpha _1^2  - \alpha _1 }
\right)R^2 } \right] = \mbox{ const}.
$$
Since solitary waves correspond to   homoclinic loops, we thus  need to state the conditions when  saddle separatrices form a loop.
Omitting the detail description \cite{skur_akust} of  phase plane of dynamical system (\ref{syst4}), let us consider the fixed saddle points and their separatrices only.
System (\ref{syst4}) has the fixed points  $O(0;0)$ and $A_{\pm}=\pm\sqrt{(1-\alpha_1)/\alpha_3}$ if $(1-\alpha_1)/\alpha_3>0$. It easy to see that the origin is a center at $\alpha_1<0$, whereas it is a saddle if $0<\alpha_1<1$ .

\begin{figure}
\begin{center}
\includegraphics[width=2.3 in, height=1.5 in]{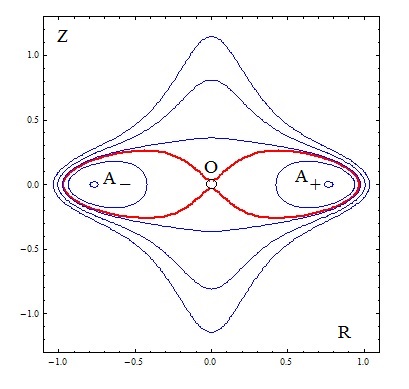} \hspace{0.2 cm}
\includegraphics[width=2.3 in, height=1.5 in]{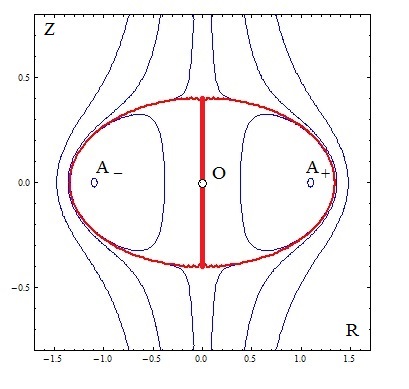}\\
a \hspace{6 cm} b
\caption{The phase portraits for  system (\ref{syst4}) at a: $\alpha_1<1$ ($D=0.9$) and b: $\alpha_1=0$ ($D=1$).}\label{fig:1}
\end{center}
\end{figure}

 Therefore, we can restrict ourself by the case when $0<\alpha_1<1$  only.  So,  the homoclinic trajectories that go through 
the origin satisfy the equation $H=0$. From this it follows the relation for $Z=Z(R)$ and the expression $s-s_0=\int dR/Z(R)$ for homoclinic loop  which can be written in the explicit form 
\begin{equation}\label{soliton}
\begin{array}{c}
\displaystyle  s - s_0  = \frac{{3 }}{2\Omega}\arcsin \left(
{\frac{{4\alpha _3 r - 3 + 4\alpha _1 }}{{\sqrt {9 - 8\alpha _1 }
}}} \right) - \frac{1}{2\Omega} \sqrt{
\frac{\alpha _1 }{1 - \alpha _1 }} \times\\ \\
\displaystyle \ln \left(
 \frac{1}{r} + \frac{{3\alpha _3  - 4\alpha _1 \alpha _3 }}{{4\left( {\alpha _1  - \alpha _1^2 } \right)}} +
 \sqrt {\left\{ {\frac{1}{r} + \frac{{3\alpha _3  - 4\alpha _1 \alpha _3 }}{{4\left( {\alpha _1  - \alpha _1^2 } \right)}}} \right\}^2  - \frac{{\alpha _3^2 \left( {9 - 8\alpha _1 } \right)}}{{16\left( {\alpha _1  - \alpha _1^2 } \right)^2 }}}
 \right)  \Biggr|_{r_0 }^r,
\end{array}
\end{equation}
where $ r=R^2$. The typical phase portrait of dynamical system (\ref{syst4}) is depicted in fig.~\ref{fig:1}a. The pair of homoclinic orbits are drawn with the  bold curves.  Solution (\ref{soliton}) corresponds to the soliton solution
with infinite support at $0<\alpha_1<1$.

When $\alpha_1$ tends to zero, the angles between separatrices of saddle point $O$ are growing. As a result,   at  $ \alpha _1 = 0 $  we obtain the degenerate phase portrait presented in fig.\ref{fig:1}b. The bold lines mark the orbits corresponding to  solitary wave  solutions with finite support (compacton). These orbits are described by the following expressions
\[
 R = U^\prime _s=\left\{ {\begin{array}{*{20}c}
   {\sqrt {\frac{3}{{2\alpha _3 }}} \sin \left( {\frac{{\Omega s}}{3}} \right),\frac{{\Omega s}}{3} \in \left[ {0;\pi } \right]}  \\
    {0,\frac{{\Omega s}}{3} \notin \left[ {0;\pi } \right]}.  \\
\end{array}} \right.
\]

This regime
can be thought as a limit state for  solution (\ref{soliton}) with infinite
support.

\subsection{Homoclinic loops in the model with cubic nonlinearity in the equation of motion for oscillating inclusions}\label{sec12}

If $\delta\not = 0$, then system (\ref{syst3}) does not reduce to the dynamical system in the plane $(R;R')$. But the first integral  for  (\ref{syst3})  can still be derived in the form
$$
I=\Omega^2\left(W-U\right)^2+\frac{\delta}{2D^2}\left(W-U\right)^4+  \alpha _3^2 R^6  + \frac{{4\alpha
_1 - 3}}{2}\alpha _3 R^4  + \left( {\alpha _1^2  - \alpha _1 }
\right)R^2 .
$$

\begin{figure}[h]
\begin{center}
\includegraphics[width=2.3 in, height=1.5 in]{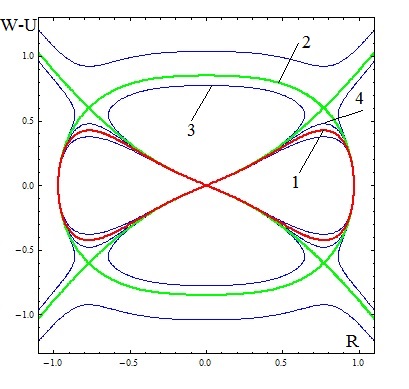}
\hspace{0.2 cm}
\includegraphics[width=2.3 in, height=1.5 in]{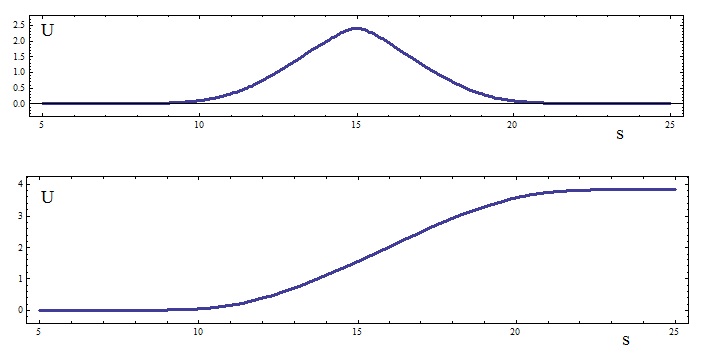}\\
%a \hspace{6 cm} b
\caption{Left: Position of level curves  $I(\delta)=0$  at different values of $\delta$. Curve 1 is plotted at $\delta=0$, curve 2 at $\delta_0$, curve 3 at $\delta=-1.5>\delta_0$, curve 4 at  $\delta=-2.7<\delta_0$.  Right: Homoclinic trajectories from the left diagram corresponding to $\delta=-2.7<\delta_0$ (upper panel) and $\delta=-1.5>\delta_0$ (lower panel). }\label{fig:2}
\end{center}
\end{figure}

Consider the position and the form of homoclinic trajectories when the parameter $\delta$ is varied. Since  homoclinic loops are described by the level curve $I=0$, we plot the set of curve $I(\delta)=0$ (fig.\ref{fig:2}). Starting from the loop 
$I(0)=0$ which coincides with the orbits of fig.\ref{fig:1}a, we see that increasing $\delta$ causes the attenuation
 of  loop's size along vertical exis. If $\delta$ decreases,  loop's size grows, but at $\delta_0$ the additional heterocycle connecting four new saddle points appears.
 The bifurcational value $\delta_0$  can be derived via  analysing the function $I(\delta)$. Namely, solving the biquadratic equation $I=0$ with respect to $W-U$, several branches of level curves are obtained. The condition of contact for two branches leads us to a qubic equation with respect to $R^2$  with zero discriminant. Then $\delta_0=-\frac{\alpha_3\Omega^4 D^2}{(\alpha_1-1)^2}$ or $\delta_0=\frac{27 \alpha_3\Omega^4 D^2}{\alpha_1^2 (9-8\alpha_1)}$.

So, if $D=0.9$,  then $\delta_0=-2.24732$. From  the figure \ref{fig:2} it follows that for $\delta<\delta_0$ the homoclinic loops are placed in the vertical quarters of the phase plane.  Note that the profiles of the resulting solitary waves are different (fig.\ref{fig:2}(right panel)), namely, at $\delta<\delta_0$ the $U$ profile looks like a bell-shape curve, whereas at $\delta>\delta_0$ it is a kink-like regime.

\subsection{The numerical scheme for the model (\ref{model})}

To model the soliton dynamics, we construct the three level
finite-difference numerical scheme for model (\ref{model}) and study its stability. 

Let us construct the numerical scheme for model (\ref{model}) in
the region $ \Sigma = \left[ {0 \le x \le L} \right] \times \left[
{0 \le t \le T} \right] $ with grid lines  $ x = ih $ and
  $
t = j\tau $, where $h$, $\tau$ are constant, $i=1...N$. Consider three level
numerical scheme. Denote  $ u = u\left( {t_{j + 1} } \right) $, $
v = u\left( {t_j } \right) $, $ q = u\left( {t_{j - 1} } \right)
$, and $ K = w\left( {t_{j + 1} } \right) $, $ G = w\left( {t_j }
\right) $, $ F = w\left( t_{j -1} \right)$. We use the following
difference approximation of derivatives
$$
\frac{{\partial ^2 u}}{{\partial t^2 }} \approx \frac{{u_i  - 2v_i
+ q_i }}{{\tau ^2 }}, \quad   \frac{{\partial ^2 u}}{{\partial x^2
}} \approx r \frac{{u_{i - 1}  - 2u_i  + u_{i + 1} }}{{h^2 }}
+ \left( {1 - r } \right)\frac{{v_{i - 1}  - 2v_i + v_{i + 1}
}}{{h^2 }},
$$
$$
\frac{{\partial u}}{{\partial x}} \approx \frac{{v_{i + 1}  - v_{i
- 1} }}{{2h}}, \quad   \frac{{\partial ^2 w}}{{\partial t^2 }}
\approx \frac{{K_i  - 2G_i  + F_i }}{{\tau ^2 }},\quad
 w - u \approx \frac{{3G_i  - F_i }}{2} - \frac{{3v_i  - q_i }}{2}\equiv \psi_i.
$$
 Thus, if $ r  \not = 0 $, then we obtain the three level implicit
 scheme:
\begin{equation}\label{num_sch}
A_i u_{i - 1}  + C_i u_i  + B_i u_{i+1}  = Y_i,\quad
 K_i  = 2G_i -
F_i  - \tau ^2 \omega ^2 \psi_i -  \tau ^2 \delta  \psi_i^3,
\end{equation}
where $\displaystyle  A_i  = B_i  = \frac{r }{{h^2 }}\varphi
$,  $\displaystyle  C_i = - 2\frac{r }{{h^2 }}\varphi -
\frac{1}{{\tau ^2 }} $, $\displaystyle \varphi=\rho^{-1}\left( {e_1  + 3e_3 \left(
{\frac{{v_{i + 1}  - v_{i - 1} }}{{2h}}} \right)^2 } \right)$,

$ \displaystyle Y_i = -\Biggl( \varphi \frac{{v_{i - 1} - 2v_i +
v_{i + 1} }}{{h^2 }}\left( 1 - r \right) + m\omega ^2 \psi_i +m\delta \psi_i^3+
\frac{{2v_i  - q_i }}{{\tau ^2 }} \Biggr)$.

\begin{figure}
\begin{center} 
\includegraphics[width=10 cm, height=5 cm]{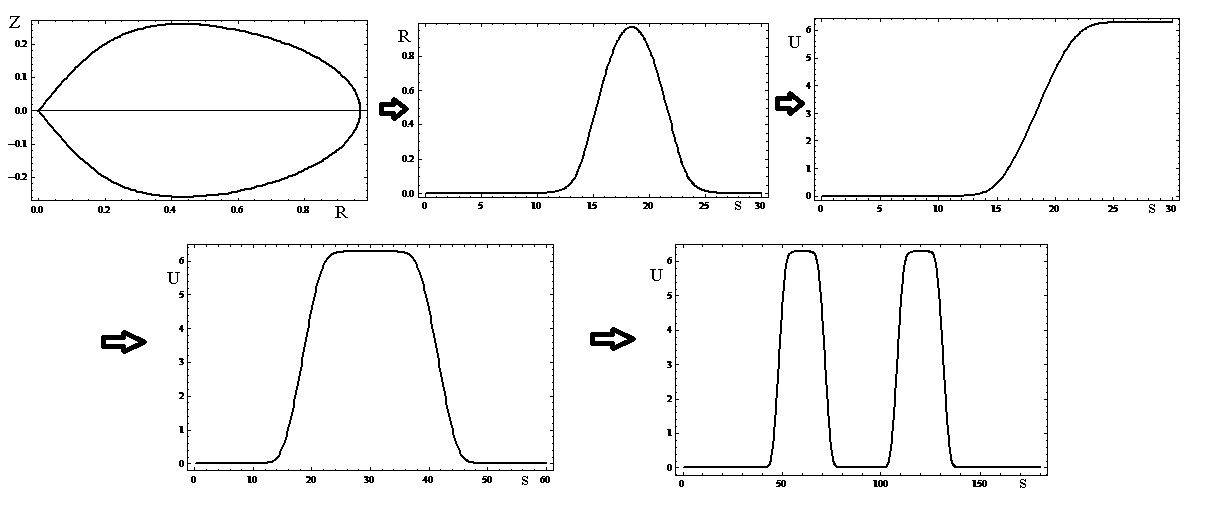}\\
\caption{The construction of initial data for numerical scheme.}\label{skur:fig3}
\end{center}
\end{figure}

System of algebraic equations (\ref{num_sch}) can be solved by the
sweep method. The necessary conditions of the sweep method
stability ($ \left| {A_i } \right| + \left| {B_i } \right| \le
\left| {C_i } \right| $ ) are fulfilled. To get the restrictions for spatial and temporal steps, the Fourier stability method, being applied to  linearized scheme (\ref{num_sch}),  is used. Let us fix the values of
the parameters $ e_1  = 1 $, $ e_3  = 0.5 $, $\rho=1$, $ \omega  =
0.9 $, $ m = 0.6 $, $ D = 0.9 $, and the parameters of the
numerical scheme $ L = 30 $, $ N = 200$, $ r = 0.3 $, $h =
L/N $, $ \tau  = 0.05 $. 

\begin{figure}[h]
\begin{center} 
\includegraphics[width=8 cm, height=6 cm]{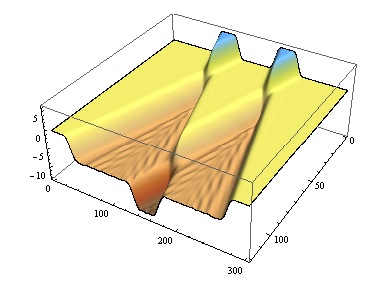}\\
\caption{ The propagation of  solitary waves at $\delta=0$ starting from the initial profile depicted in the  figure \ref{skur:fig3}.}\label{skur:fig4}
\end{center}
\end{figure}

Let us construct the initial data $v_i$, $q_i$, $G_i$, $F_i$ for numerical simulation on the base of solitary waves. To do this, we integrate dynamical system (\ref{syst4}) with initial data $R(0)=10^{-8}$, $Z(0)=0$, $s \in [0;L]$  and choose the right  homoclinic loop in the phase portrait (fig.\ref{fig:1}). Then the profiles of $W(s)$, $U(s)$, and $R(s)$ can be derived. Joining the proper arrays, we can build the profile in the form of  arch:
$$
v=U(ih)\cup U(L-ih),\, q=u(x+\tau D)=\left[U(ih)+\tau D R(ih)\right]\cup \left[U(L-ih)+\tau D R(L-ih)\right].
$$
The arrays $G$ and $F$ are formed in similar manner. 
%Note that the length of $v$ ($q$, $G$, $F$) is now $K=2N$. 
Combining two arches and continuing the steady solutions 
%adding 200 boundary points 
at the ends of graph, we get more complicated profile.% Now $K=4N+400$.
The sequence of steps for profile construction is depicted in figure \ref{skur:fig3} in detail. We apply the fixed boundary conditions, i.e. $u(x=0,t)=v_1$,   $u(x=Kh,t)=v_K$, where $K$ is the length of an array.

Starting from the two-arch initial data, we see (fig.\ref{skur:fig4}) that solitary waves move to each other, vanish during approaching, and appear with negative amplitude and shift of phases. After collision in the zones between  waves some ripples are revealed.  Secondary collision of waves are watched also. Note that the simulation of compacton solutions displays similar properties. 

\begin{figure}[h]
\begin{center} 
\includegraphics[width=12 cm, height=4.5 cm]{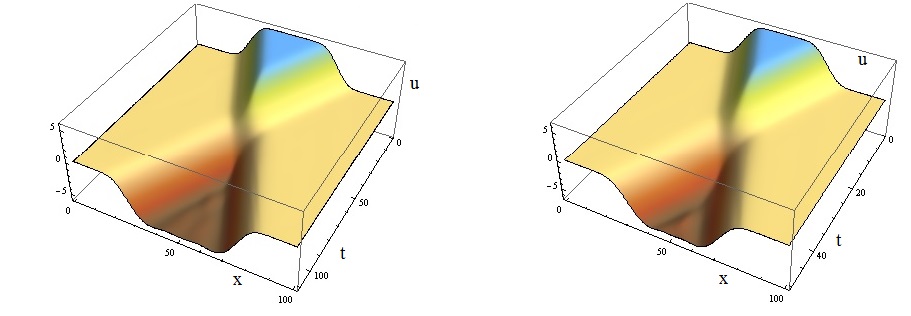}\\
\caption{Solitary waves dynamics at $\delta=-0.2$ and $\delta=1.3$.}\label{skur:fig5}
\end{center}
\end{figure}

\begin{figure}[h]
\begin{center} 
\includegraphics[width=6 cm, height=4.5 cm]{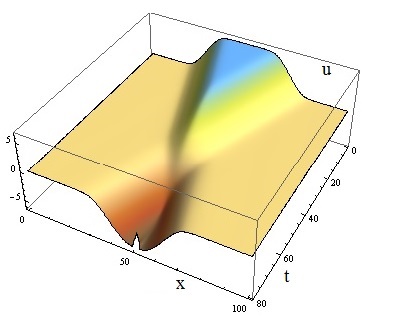}\hspace{0.5 cm}
\includegraphics[width=6 cm, height=4.2 cm]{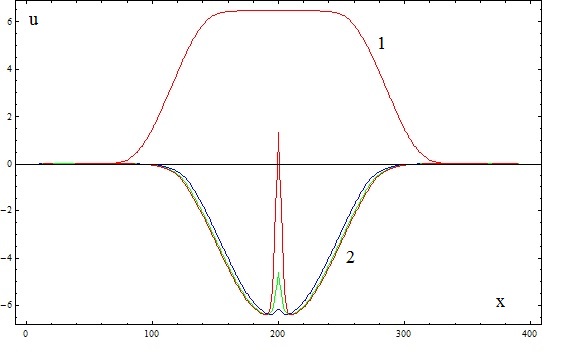}\\
\caption{Solitary waves dynamics at $\delta=-0.3$. In the right panel the curve 1 is the initial profile, whereas other curves 2 are the forms of solutions just before blowup.}\label{skur:fig6}
\end{center}
\end{figure}

Propagation of solitary waves at $\delta \not =0$ depends on the sign of $\delta$. Analysing the  diagrams of figures \ref{skur:fig5}, we see that for $\delta>0$ the collision of waves is similar to the collision at  $\delta=0$. Behaviour of waves after interaction does not change essentially  when  $\delta<0$ and close to zero. But at $\delta=-0.3$  after collision the amplitude of solution is increasing in the place of soliton's intersection (fig.\ref{skur:fig6}) and  after a while   the solution  is  destroyed. 

 This suggests that we encounter  the unstable  interaction of solitary waves or the numerical scheme we used possesses spurious solutions. But if we take half spatial step and increase the scheme  parameter $r$ up to 0.8, the scenario of solitary waves collision is not changed qualitatively. Therefore, the assertion on the unstable nature of collision is more preferable.

\section{Conclusions}

So, in this report, we have presented the novel  nonlinear generalization of model for media with oscillating inclusions. It is important for application  that the new parameters $e_3$ and $\delta$ have clear physical meaning.  As it was shown in subsections \ref{sec11} and \ref{sec12},  this model possesses the solitary waves of different  types including compactons. 
We have considered the conditions of their existence and bifurcations when the parameters of nonlinearity were varied. The exact solutions describing the solitary waves with both   unbounded and compact support were derived.
Analysing the expression (\ref{soliton}) and Fig.~\ref{fig:2}a,  we should emphasize  that the characteristics of  solitary waves crucially depend on the dynamics of oscillating inclusions. 
 To study the properties of solitary waves and their interactions, we have proposed  the effective numerical scheme based on the finite difference approximation of the continious model.    In particular, we have found out the conditions when  the stable propagation of solitons and their collisions are observed. It turned out that the nonlinearity of the oscillating dynamics describing by $\Phi$  affects not only the form of solitary waves but their stability properties in  collisions (Fig.~\ref{skur:fig6}).
Finally, there are solitary waves moving without preserving of their selfsimilar  shapes.   While the numerical results concerning  the stable properties of single solitary wave can be confirmed by analytical treatments \cite{vsanCNSNS}, the studies of wave collisions require mostly the application of improved numerical schemes.
The results presented above can be useful for modelling the behaviour of  complex media in vibrational fields, for instance when we deal with the  intensification of extracting oil and natural gas \cite{Nikol}.

%From the results presented above it follows that the model of media with oscillating inclusions possesses the solitary waves of %different  types. These solutions help us to derive the effective numerical scheme and study the dynamics of more complicated %regimes such as multi-collisions of solitonic waves. Within the framework of scheme  we have seen that the propagation of solitary %waves essentially depend on dynamics of oscillating inclusions. In particular, the stable solitons and their collisions are observed. %We have computed the stable single impulse solutions which manifest unstability after interaction. Finally, there are solitary waves %moving without preserving of their selfsimilar  shapes.   While the numerical results concerning  the stable properties of single %solitary wave can be confirmed by analytical treatments \cite{vsanCNSNS}, the studies of wave collisions require mostly the %application of improved numerical schemes.

%\section*{Acknowledgments}
%The authors have been supported by ..., etc.
%
%Inclusion of bibliography
%
%\halfnormalsize
\bibliographystyle{acm} 	%% The bibliography style file 'acm.bst' must be available in the folder.
\bibliography{DSTA15_DAN_SKUR} 	%% A file dsta-2015.bib must be available in the folder.
%							%% Use \cite{} to inclue a reference in the text.
%
% AUTHOR STAMPS AT THE END MUST BE INSERTED
%
% Put author stamps in the order that they appear under the title on first page.
% Use more/less \stamp commands to add/remove any author stamps.
% At least one author has to be marked as the speaker by adding the note in last parameter of \stamp command:
% "The author gave a presentation of this paper during one of the conference sessions." (see an example below)
%

%
%end
%
\end{document}